\documentclass[twocolumn,showpacs,aps,prl,floatfix,superscriptaddress]{revtex4}
\usepackage{amsmath,amssymb,eucal}
\usepackage{graphicx}
\usepackage{float}
\usepackage{multirow}

\begin{document}

\title{Growth Inside a Corner: The Limiting Interface Shape}

\author{Jason Olejarz}
\affiliation{Center for Polymer Studies, and Department of Physics, Boston University, Boston, MA 02215, USA}
\author{P.~L.~Krapivsky}
\affiliation{Department of Physics, Boston University, Boston, MA 02215, USA}
\author{S.~Redner}
\affiliation{Center for Polymer Studies, and Department of Physics, Boston University, Boston, MA 02215, USA}
\author{K.~Mallick}
\affiliation{Institut de Physique Th\'{e}orique CEA, IPhT, F-91191 Gif-sur-Yvette, France}

\begin{abstract}
  We investigate the growth of a crystal that is built by depositing cubes
  inside a corner.  The interface of this crystal approaches a deterministic
  growing limiting shape in the long-time limit.  Building on known results
  for the corresponding two-dimensional system and accounting for basic
  three-dimensional symmetries, we conjecture a governing equation for the
  evolution of the interface profile.  We solve this equation analytically
  and find excellent agreement with simulations of the growth process.  We
  also present a generalization to arbitrary spatial dimension.
\end{abstract}


\maketitle

Growing interfaces constitute a venerable subject, but the proper continuum
framework to account for this growth was developed not so long
ago~\cite{KPZ}.  A detailed and beautiful description of fluctuations of {\em
  one-dimensional} growing interfaces has been proposed~\cite{HZ95,KK10},
culminating in a recent solution of the KPZ equation \cite{KPZ_rev}.  For
real applications, two-dimensional growing interfaces are much more
important, but their governing stochastic continuum equations~\cite{KPZ}
remain unsolved.  Nevertheless, the analysis of two-dimensional growing
interfaces is not hopeless.  Indeed, although interface fluctuations have
attracted the most attention, they become less important as the interface
grows.  The limiting shape --- the average interface profile in the long-time
limit --- is the more primal characteristic.

If growth begins from a flat substrate, the interface advances at a constant
average speed, so only fluctuations matter.  In numerous applications,
however, the limiting shapes are curved and are known only in rare cases.
One such example is the 2+1 dimensional Gates-Westcott model for vicinal
interfaces, which was solved by a free-fermion mapping~\cite{Prahofer}.  This
growth process exhibits logarithmic height correlations and therefore does
not belong to the strong-coupling KPZ universality class.  Average interface
profiles are also known for certain anisotropic 2+1 dimensional growth
models~\cite{RajeshDhar,Borodin}.  However, even for the most basic isotropic
growth models limiting shapes are not known.  For example, for the
two-dimensional Eden-Richardson model~\cite{eden} the limiting shape is
unknown, although the statistics of its fluctuations are understood (and
belong to the KPZ universality class).

\begin{figure}[ht!]
\begin{center}
\raisebox{1.0in}{\includegraphics*[width=0.12\textwidth]{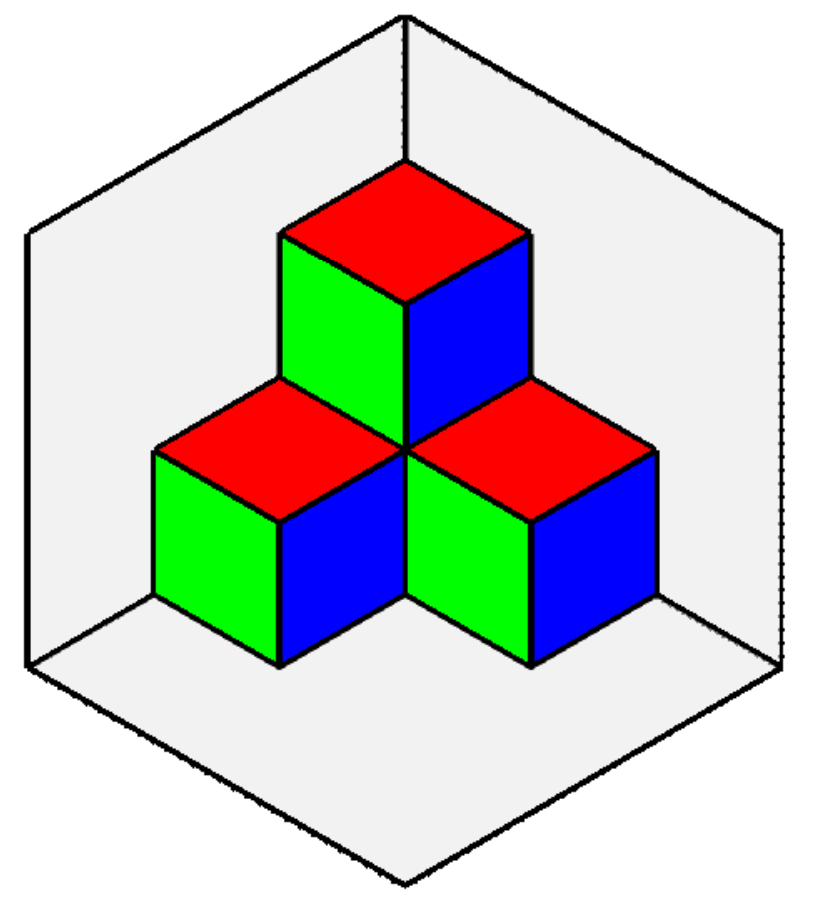}}\includegraphics*[width=0.325\textwidth]{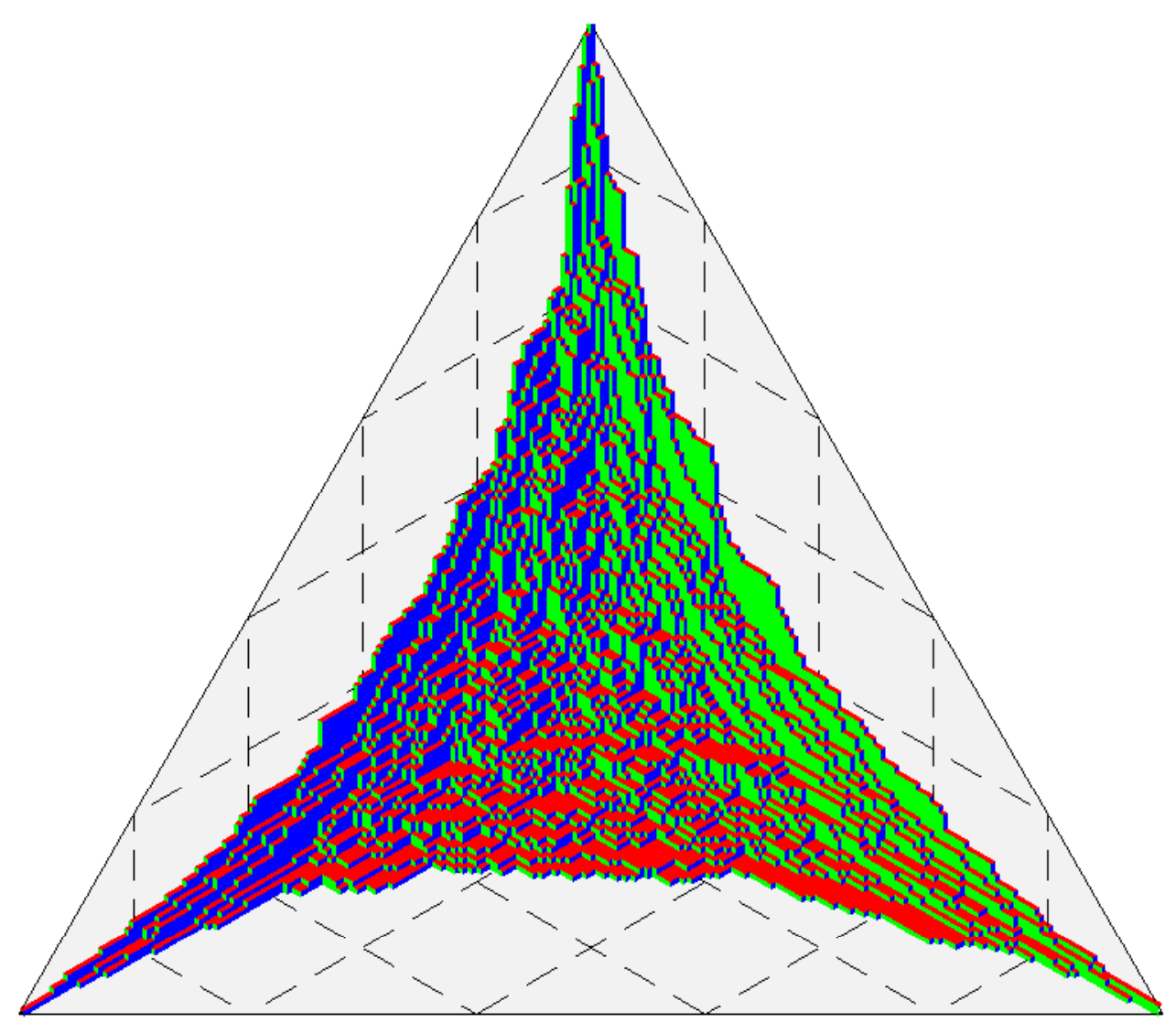}
\caption{\small (color online) Upper left: 3d crystal of volume 4.  The next
  elemental cube can be deposited at one of 6 inner corners.  Right: Crystal
  at $t=140$.}
\label{fig:interface}
  \end{center}
\end{figure}

Here we investigate the limiting shape of a crystal that grows inside a
corner.  This process can be defined in arbitrary dimension and on any
lattice (with an appropriately defined `corner').  We specifically consider a
cubic lattice, where the corner is the initially empty positive octant.
Starting at $t=0$, elemental cubes are deposited at unit rate onto inner
corners (Fig.~\ref{fig:interface}).  Initially, there is one inner corner and
thus one place where a cube can deposit.  After this first event, there are
three available inner corners that can accommodate the next cube.  The
interface shape becomes smoother as it grows and ultimately approaches a
deterministic limiting shape.

The corner growth model admits a dual interpretation as the melting of a
three-dimensional cubic crystal by erosion from the corner.  There is also a
magnetic interpretation in which plus spins are initially assigned to each
site inside the corner and minus spins to exterior sites, with the spins
endowed with zero-temperature Glauber spin-flip dynamics~\cite{G63} in a weak
negative magnetic field.  This dynamics allows only plus spins at inner
corners to flip and thus is isomorphic to the corner melting problem.  The
magnetic interpretation naturally suggests considering the system in zero
magnetic field, which results in a growing interface whose characteristic
scale grows diffusively rather than ballistically.  Other modifications
involve changing the initial condition; e.g., depositing the cubes onto a
planar substrate (the `hypercube stacking model' \cite{step}) leads to a
trivial limiting shape but is better-suited to studying non-trivial height
fluctuations.

In what follows, we use the language of deposition; most importantly, we
allow only deposition events and no evaporation.  Growth inside a
two-dimensional corner is well understood by mapping the corner growth
process onto the one-dimensional asymmetric exclusion process~\cite{rost};
fluctuations in this limiting shape have also been computed~\cite{BDJ,J00}.
In three dimensions, the corner growth model can be mapped into an infinite
set of coupled exclusion processes in the plane, also known as the `zigzag
model' \cite{Tamm,details}.  Unfortunately, no exact solutions are known for
such planar interacting particle processes.
 
Here we focus on the limiting shape in three (and higher) dimensional
corners.  Our analysis relies heavily on insights gleaned from the limiting
two-dimensional corner interface shape~\cite{rost}.  In two dimensions this
limiting shape $y(x;t)$ evolves according to the equation of
motion~\cite{Liggett,spohn,karma}
\begin{equation}
\label{eqn:yt-2d}
y_t=\frac{y_x}{y_x-1}~,
\end{equation}
from which the interface profile was found to be \cite{rost}
\begin{equation}
\label{eqn:shape-2d}
\sqrt{x}+\sqrt{y}=\sqrt{t}\,.
\end{equation}
This parabolic shape \eqref{eqn:shape-2d} describes the non-trivial part of
the interface where $0\leq x, y\leq t$.  Outside this region, the original
boundary is undisturbed.

Two properties severely constrain the form of possible evolution equations
for growth inside a three-dimensional corner: (a) The governing equation for
the interface $z(x,y;t)$ must reduce to the two-dimensional form
\eqref{eqn:yt-2d} on the boundaries $x=0$ or $y=0$; (b) The equation must be
invariant under the interchange of any coordinate pair.

Analogously to Eq.~\eqref{eqn:yt-2d}, we seek a three-dimensional evolution
equation of the form $z_t = F(z_x, z_y)$ that involves only first derivatives
(higher-order derivatives are asymptotically negligible).  The simplest guess
is the product
$z_t=\left[{z_x}/({z_x-1})\right]\,\left[{z_y}/({z_y-1})\right]$.  This
equation reduces to \eqref{eqn:yt-2d} on the boundaries $x=0$, where
$z_x=-\infty$, and $y=0$, where $z_y=-\infty$.  The product ansatz is also
invariant under the exchange $x\leftrightarrow y$ but {\em not} under the
exchanges $x\leftrightarrow z$ or $y\leftrightarrow z$ and therefore is
wrong.

By extensive trial and error, we found that
\begin{equation}
\label{zt-3d-final}
z_t =\frac{z_x}{z_x-1}\, \frac{z_y}{z_y-1} \left[1-\frac{1}{z_x+z_y}\right]
\end{equation}
satisfies the necessary coordinate interchange invariances.  These
constraints severely limit the form of the evolution equation.  For example,
if we seek a multiplicative correction factor to the product form in
\eqref{zt-3d-final} as the Laurent series
$\sum_{-\infty}^{\infty}\lambda_n(z_x+z_y)^{-n}$, coordinate interchange
invariance gives $\lambda_0=1, \lambda_1=-1$, while all other amplitudes
vanish~\cite{details}.  Thus Eq.~\eqref{zt-3d-final} is the only invariant
choice among the family of evolutionary equations parameterized by
$\lambda_n$.

We also found one other elemental evolution equation of the form $z_t =
F(z_x, z_y)$ that satisfies coordinate interchange invariance; this form is
unique if we again seek corrections as a Laurent series representation.  This
second solution is obtained by replacing the factor in the square brackets in
\eqref{zt-3d-final} with $\left[1 + ({z_xz_y -z_x - z_y})^{-1}\right]$.  This
equation, which can be re-written more elegantly as
\begin{equation}
\label{grow_3}
\frac{1}{z_t} = 1 - \frac{1}{z_x} - \frac{1}{z_y}\,,
\end{equation}
and Eq.~\eqref{zt-3d-final} are two functionally independent
three-dimensional evolution equations that satisfy coordinate interchange
invariance.  We believe, but cannot prove, that other elemental evolution
equations do not exist.

\begin{figure}[ht!]
\begin{center}
\includegraphics*[width=0.3\textwidth]{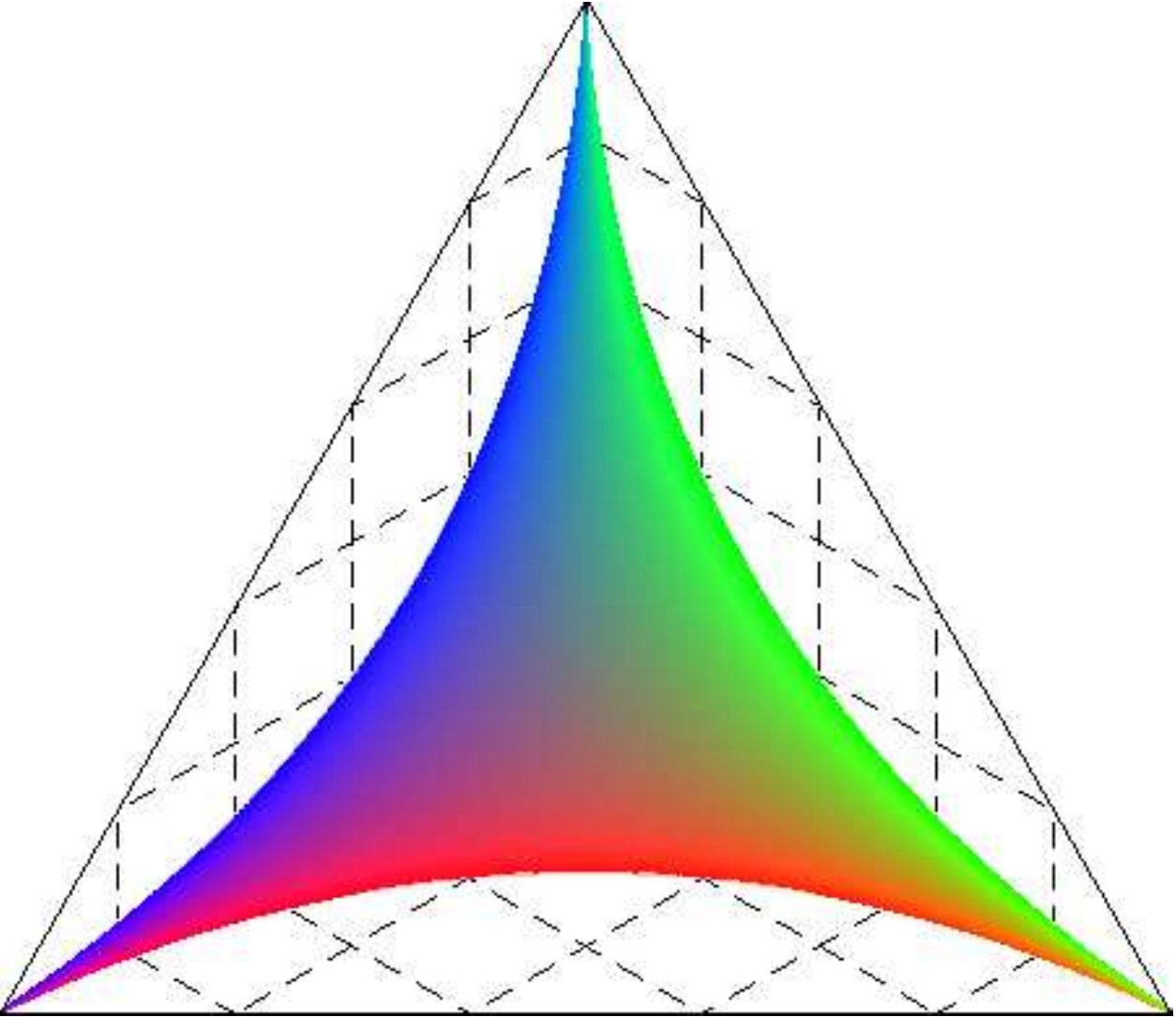}
\caption{\small (color online) The interface
  (\ref{eqn:3d_growth_limit_shape}).}
\label{fig:parametric_plot}
  \end{center}
\end{figure}

Our conjecture is that \eqref{zt-3d-final} is the correct evolution equation.
Evidence in favor of this statement also comes from the excellent agreement
with simulation data.  For this comparison, we solve Eq.~\eqref{zt-3d-final}
by the method of characteristics.  Starting from an empty corner, we
find~\cite{details} that the interface profile admits the following
parametric representation (Fig.~\ref{fig:parametric_plot})
\begin{equation}
\frac{x}{t}=A(q,r), \quad \frac{y}{t}=B(q,r), \quad \frac{z}{t}=C(q,r)
\label{eqn:3d_growth_limit_shape}
\end{equation}
where
\begin{align*}
A=&\frac{1}{(q-1)^2}\frac{r}{r-1}\left[1-\frac{1}{q+r}\right]-\frac{q}{q-1}\frac{r}{r-1}\frac{1}{(q+r)^2}~, \\
 B=&\frac{q}{q-1}\frac{1}{(r-1)^2}\left[1-\frac{1}{q+r}\right]-\frac{q}{q-1}\frac{r}{r-1}\frac{1}{(q+r)^2}~, \\
 C=&\frac{q}{q-1}\frac{r}{r-1}\left[1-\frac{1}{q+r}\right]\left[1+\frac{1}{q-1}+\frac{1}{r-1}\right] \\ &-\frac{q}{q-1}\frac{r}{r-1}\frac{1}{q+r}~,
\end{align*}
with $q=z_x$, $r=z_y$ and $-\infty<q,r\leq 0$.  As a consistency check, note
that for $r=-\infty$, we have $x/t=(q-1)^{-2}$, $y/t=0$, and
$z/t=q^2(q-1)^{-2}$.  Eliminating $q$, we get $\sqrt{x}+\sqrt{z}=\sqrt{t}$,
thereby recovering Eq.~\eqref{eqn:shape-2d} for the intersection of the
interface \eqref{eqn:3d_growth_limit_shape} with the $y=0$ plane.

\begin{figure}[ht!]
\begin{center}
  \includegraphics*[width=0.355\textwidth]{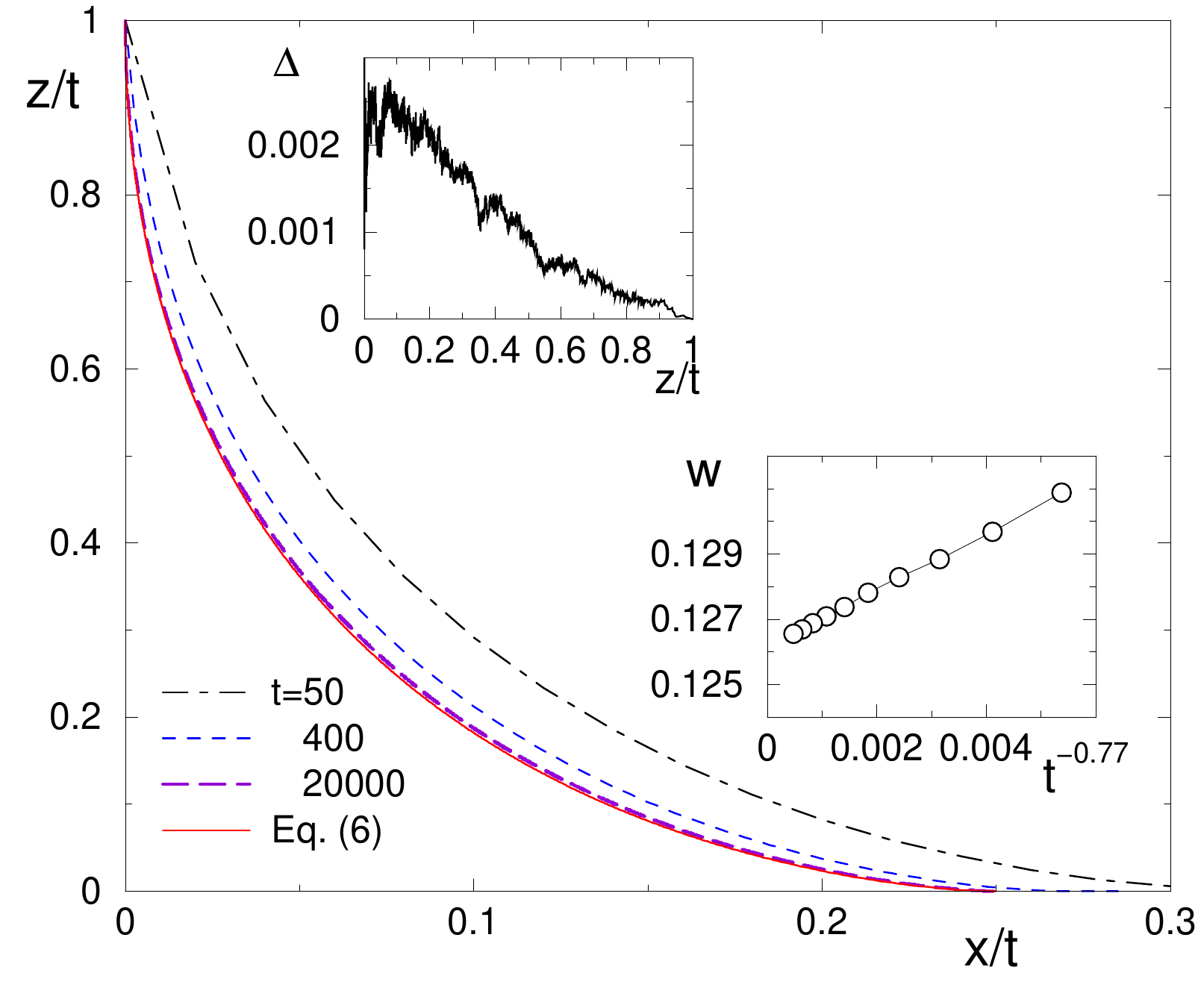}
  \caption{\small (color online) Scaling plot of the surface profile $z/t$
    versus $x/t$ along the diagonal $x=y$ at different times.  Upper-left
    inset: the difference $\Delta$ between the simulated values of the left-
    and right-sides of \eqref{eqn:diag}.  Lower-right inset: convergence of
    the diagonal interface speed versus $t^{-0.77}$.}
\label{fig:diag}
  \end{center}
\end{figure}

It seems impossible to eliminate the parameters $(q,r)$ from
Eq.~\eqref{eqn:3d_growth_limit_shape} and obtain a closed-form representation
of the interface in terms of $x,y,z$ and $t$ as in the two-dimensional case.
However, the intersections of the interface \eqref{eqn:3d_growth_limit_shape}
with certain planes admit simplified descriptions.  For example, for the
plane $x=y$, corresponding to $q=r$, we obtain
\begin{equation}
\frac{x}{t}=\frac{1}{2}\frac{z}{t}-\frac{3}{4}\left(\frac{z}{t}\right)^{2/3}+\frac{1}{4}~,
\label{eqn:diag}
\end{equation}
which agrees well with simulations (Fig.~\ref{fig:diag}).

Two additional tests suggest that the conjectured evolution equation
\eqref{zt-3d-final} and its solution \eqref{eqn:3d_growth_limit_shape}
describe corner growth accurately.  Consider first the advance of the
interface along the ray $x=y=z$.  From \eqref{eqn:3d_growth_limit_shape}, the
position of this point is given by \cite{middle}
\begin{equation}
  x=y=z=wt, \qquad w =\tfrac{1}{8}\,.
\label{middle_point}
\end{equation}
Numerically, we measure $w\approx 0.1261(2)$, which agrees with our
prediction $w=0.125$ to within $0.9\%$.  As a second test, we compute the
total volume $V$ beneath the growing interface at time $t$.  Since the linear
dimension of the interface grows linearly with time, $V=vt^3$.  To determine
the amplitude $v$, we use the parametric solution
\eqref{eqn:3d_growth_limit_shape} and change from the physical variables
$(x,y)$ to the parametric coordinates $(q,r)$, from which the amplitude $v$
reduces to the integral
\[
v=\int_{-\infty}^0\int_{-\infty}^0 dq\;dr\;C(q,r)\;\frac{\partial(A,B)}{\partial(q,r)}~.
\]
We compute the Jacobian $\frac{\partial(A,B)}{\partial(q,r)}$ and the
integral using \emph{Mathematica} and find
\begin{equation}
\label{volume}
v=\frac{3\pi^2}{2^{11}}=0.014457\ldots
\end{equation}
Numerically, we measure $v\approx 0.01472(3)$, which is within $1.8\%$ of our
prediction.

While Eq.~\eqref{zt-3d-final} accurately describes the corner interface,
small discrepancies between our measurements of the coefficients $w$ and $v$,
and their predicted values \eqref{middle_point}--\eqref{volume} persist.  The
alternative elemental evolution equation \eqref{grow_3} leads to the
interface profile
\begin{equation}
\label{wrong:shape-3d}
\sqrt{x} + \sqrt{y} + \sqrt{z}=\sqrt{t}\,,
\end{equation}
which is the natural generalization of Eq.~\eqref{eqn:shape-2d}. The
corresponding values $w=\frac{1}{9}$ and $v=\frac{1}{90}$ that arise from
this profile substantially disagree with simulation results, suggesting that
\eqref{grow_3} is wrong.

From the elemental equations \eqref{zt-3d-final} and \eqref{grow_3}, we can
also form two distinct one-parameter families of invariant
equations~\cite{details}; an additive family
\begin{subequations}
\label{alt}
\begin{equation}
z_t =\frac{z_x}{z_x\!-\!1}\, \frac{z_y}{z_y\!-\!1} \left[1\!-\!\frac{1+c}{z_x+z_y} \!-\! \frac{c}{z_xz_y \!-\!z_x \!-\! z_y}\right]\!,
\end{equation}
and a multiplicative family 
\begin{equation}
\label{multiplicative}
z_t =
\left[\frac{1-\frac{1}{z_x+z_y}}{\big(1-\frac{1}{z_x}\big) \big(1-\frac{1}{z_y}\big)}\right]^{1+c}
\left[1 - \frac{1}{z_x} - \frac{1}{z_y}\right]^c\!,
\end{equation}
\end{subequations}
where for both families the limit $c=0$ reduces to \eqref{zt-3d-final} and
the limit $c=-1$ reduces to \eqref{grow_3}.  For the multiplicative class of
evolution equations~\eqref{multiplicative}, the choice $c\approx 0.074$
provides the best fit for the simulated value of $v$ \cite{vel}.  Similarly,
for the additive class of equations, the optimal mixing parameter is
$c\approx 0.079$.  However, a phenomenon as minimalist as corner interface
growth should be described by a simple equation that does not contain an
anomalously small mixing parameter.  This aesthetic consideration, in
conjunction with our numerical results, suggest that Eq.~\eqref{zt-3d-final}
describes corner interface evolution.

The small discrepancies between our simulation results and the predictions
that follow from Eq.~\eqref{zt-3d-final} (see the insets to
Fig.~\ref{fig:diag}) suggest that the approach to the asymptotic state is
slow.  A similarly slow convergence to asymptotic behavior occurs in various
well-understood one-dimensional growth models (see
e.g. Refs.~\cite{farnudi,ferrari}).  For example, for 1+1 dimensional corner
growth, the intersection of the interface with the $(1,1)$ direction evolves
according to~\cite{BDJ,J00,KK10}
\begin{equation}
\label{fluct_2}
x(t) = \frac{t}{4} + t^{1/3}\, \xi\,,
\end{equation}
where $\xi$ is a stationary random variable with $\langle \xi\rangle >0$.
Thus averaging over many realizations gives an effective velocity
$w_\text{eff}-\frac{1}{4} \sim t^{-2/3}$.

For growth inside a three-dimensional corner, we therefore anticipate that
$w_\text{eff} - \frac{1}{8}\sim t^{-\alpha}$, with a still-unknown exponent
$\alpha$.  Very extensive simulations for flat interfaces in 2+1 dimensions
indicate that $\alpha$ is close to 0.77~\cite{KrugMeakin,reis,odor}.  On the
other hand, extrapolation from our simulations for $t\alt 20000$ suggests
that $\alpha\approx 0.74$.  This difference in exponent estimates suggests
that $t=20000$ is still outside the long-time regime for growth inside a
three-dimensional corner.  This slow approach to asymptotic behavior could be
the source of the discrepancy between our simulation results and the
theoretical prediction \eqref{zt-3d-final} for the interface profile.


Our argument for the form of the evolution equation can be generalized to
higher dimensions.  Applying coordinate interchange invariance and related
symmetry considerations, we conjecture that in $d$ dimensions the height
$h(x_1,\ldots,x_{d-1}; t)$ satisfies
\begin{equation}
\label{grow_d}
h_t=\prod_{1\leq i_1<\ldots<i_p\leq d-1}\left(1-\frac{1}{h_{i_1}+\ldots+h_{i_p}}\right)^{(-1)^p}
\end{equation}
where $h_i\equiv \frac{\partial h}{\partial x_i}$.  These equations are again
solvable using the method of characteristics \cite{details}.

We emphasize that computing the limiting shape --- the primary characteristic
of the interface --- represents only a first step to understanding its
properties.  One challenging problem, given that interface fluctuations are
unknown even for flat interfaces, is to generalize Eq.~\eqref{fluct_2} to
account for fluctuations of an interface that grows at a three-dimensional
corner.  Also of interest are height-height correlations at different
locations and different times.  In 1+1 dimensions, these correlations decay
slowly along the characteristic curves of the evolution equation
\cite{Borodin,Ferrari}.  Whether similar behavior occurs in 2+1
dimensional corner growth is unknown.

Fluctuations of integral characteristics of the interface, such as the
crystal volume, may be more tractable and give rise to new phenomena.
Consider, for example, the total number of sites of various fixed degrees
(number of adjacent vertices).  Sites of degree 3, in particular, can be
categorized as either inner or outer corners.  The number of inner corners
grows as $N_\text{in}=\frac{dV}{dt}=3vt^2$, with $v=3\pi^2/2^{11}$ to
leading order.  One might anticipate the same asymptotic growth for outer
corners, but simulations indicate that the latter grows slightly faster
\cite{details}:
\begin{equation}
\label{ratio}
{N_\text{out}}/{N_\text{in}}\approx 1.04~.
\end{equation}
Note that in two dimensions $N_\text{in}-N_\text{out}=1$.  For Ising corner
growth in three dimensions, $N_\text{in}-N_\text{out}$ is also positive and
grows with time as $t^{1/2}$.  This makes the behavior in \eqref{ratio} quite
puzzling.

The other major challenges are to generalize from strict corner growth to
Ising growth, where adsorption at inner corners and desorption from outer
corners occur with equal rates, and to equilibrium interfaces, where the
desorption rate exceeds the adsorption rate.  The corresponding equilibrium
shape has been determined both in two \cite{Temp} and three dimensions
\cite{CK_2001}, and its shape fluctuations have also been studied~\cite{FS}.
In analogy with the conjectured evolution equations \eqref{grow_d} for corner
growth, there may also exist an exact generalization of equilibrium limiting
shapes \cite{Temp,CK_2001} in higher dimensions.

\smallskip We thank A. Borodin and H. Spohn for useful correspondence. JO and
SR gratefully acknowledge financial support from NSF Grant No.\ DMR-0906504.


\begin{thebibliography}{99}

\bibitem{KPZ} M. Kardar, G. Parisi, and Y.-C. Zhang, Phys.\ Rev.\ Lett.\ {\bf
    56}, 889 (1986).

\bibitem{HZ95} T. Halpin-Healey and Y.-C. Zhang, Phys.\ Rep.\ {\bf 254}, 215
  (1995).

\bibitem{KK10} 
   T. Kriecherbauer and J. Krug, J. Phys.\ A {\bf 43}, 403001 (2010).

\bibitem{KPZ_rev} 
    For a review, see I. Corwin, arXiv:1106.1596.

\bibitem{Prahofer} 
    M. Pr\"ahofer and H. Spohn, J. Stat.\ Phys.\ {\bf 88}, 999 (1997). 

\bibitem{RajeshDhar} 
    R. Rajesh and D. Dhar,  Phys.\ Rev.\ Lett.\ {\bf 81}, 1646 (1998).

\bibitem{Borodin} P. Ferrari and A. Borodin, arXiv:0804.3035.

\bibitem{eden} 
    D. Richardson, Proc.\ Camb.\ Phil.\ Soc.\ {\bf 74}, 515 (1973); 
    H. Kesten, Lecture Notes in Math.\ {\bf 1180} 125 (Springer, Berlin, 1986).

\bibitem{G63} R.~J.~Glauber, J. Math.\ Phys.\ {\bf 4}, 294 (1963).

\bibitem{step} 
     M. Plischke, Z. R\'acz, and D. Liu, Phys.\ Rev.\ B {\bf 35}, 3485 (1987); 
     B. M. Forrest and L.-H. Tang,  Phys.\ Rev.\ Lett.\ {\bf 64}, 1405 (1990).

\bibitem{rost} H.~Rost,  Prob.\ Theor.\ Relat.\ Fields {\bf 58}, 41 (1981).

\bibitem{BDJ} J. Baik, P. A. Deift, and K. Johansson, J. Amer.\ Math.\ Soc.\
  {\bf 12}, 1119 (1999).

\bibitem{J00} K. Johansson, Commun.\ Math.\ Phys.\ {\bf 209}, 437 (2000).

\bibitem{Tamm} 
     M. Tamm, S. Nechaev, and S. N.  Majumdar, J. Phys.\ A {\bf 44}, 012002 (2011). 

\bibitem{details} 
     J. Olejarz, P. L. Krapivsky, S. Redner, and K. Mallick, in preparation.



\bibitem{Liggett}
        T. M. Liggett, {\it Interacting Particle Systems} (Springer, New York, 1985).
         
\bibitem{spohn} H. Spohn, {\it Large Scale Dynamics of Interacting Particles}
  (Springer, Berlin, 1991).

\bibitem{karma}
        A.~Karma and A.~E.~Lobkovsky, Phys.\ Rev.\ E {\bf 71}, 036114 (2005).

\bibitem{middle} Eqeuation~\eqref{middle_point} follows from
        \eqref{eqn:3d_growth_limit_shape} by setting $q=r=-1$. One can also
        derive \eqref{middle_point} directly from \eqref{zt-3d-final} without
        using \eqref{eqn:3d_growth_limit_shape} for the
        interface shape; it suffices to note that by symmetry $z_x=z_y=-1$ at
        the midpoint.

\bibitem{vel} Generally for Eq.~\eqref{multiplicative}, one finds
  $v=9^c/8^{1+c}$, which is best fit to the simulation data by choosing
  $c\approx 0.074$; similarly for equations from the additive class
  $8v=1+\frac{c}{9}$.

\bibitem{farnudi} B. Farnudi and D. D. Vvedensky, Phys.\ Rev.\ E {\bf 83},
  020103(R) (2011).

\bibitem{ferrari} P. L. Ferrari and R. Frings,  J. Stat.\ Phys.\ {\bf 144}, 1123 (2011).

\bibitem{KrugMeakin} 
    J. Krug and P. Meakin, J. Phys.\ A {\bf 23}, L987 (1990).

\bibitem{reis} F. D. A. Aar\~{a}o Reis, Phys.\ Rev.\ E {\bf 69}, 021610
  (2004).

\bibitem{odor} G. \'{O}dor, B. Liedke, and K.-H. Heinig, Phys.\ Rev.\ E {\bf
    79}, 021125 (2009).

\bibitem{Ferrari} 
     P. Ferrari, J. Stat.\ Mech.\ P07022 (2008). 


\bibitem{Temp} H. Temperley, Proc.\ Cambridge Philos.\ Soc.\ {\bf 48}, 683
     (1952); A. M. Vershik and S. V. Kerov, Funct.\ Anal.\ Appl.\ {\bf 19},
     21 (1985); J.-P.~Marchand and Ph.~A.~Martin, J. Stat.\ Phys.\ {\bf 44}, 491 (1986).

\bibitem{CK_2001} R. Cerf and R. Kenyon, Commun.\ Math.\ Phys.\ {\bf 222},
     147 (2001); A. Okounkov and N. Reshetikhin, J. Amer.\ Math.\ Soc.\ {\bf
       16}, 581 (2003).

\bibitem{FS} P. L. Ferrari and H. Spohn, J. Stat.\ Phys.\ {\bf 113}, 1
       (2003); P. L. Ferrari, Ph.D. Thesis, Munich University of Technology
       (2004).

                      
\end{thebibliography}
\end{document}